# Nanomagnetic Boolean Logic – The Tempered (and Realistic) Vision

**Supriyo Bandyopadhyay[1], Fellow, IEEE**

[1]Department of Electrical and Computer Engineering, Virginia Commonwealth University, Richmond, VA 23284, USA

Corresponding author: S. Bandyopadhyay (e-mail: sbandy@vcu.edu).

This work is supported in part by the US National Science Foundation under grants CCF-2006843 and CCF-2001255

**ABSTRACT** The idea of nanomagnetic Boolean logic was advanced more than two decades ago. It envisaged the use of nanomagnets with two stable magnetization orientations as the primitive binary switch for implementing logic gates and ultimately combinational/sequential circuits. Enthusiastic proclamations of how nanomagnetic logic will eclipse traditional (transistor-based) logic circuits proliferated the applied physics literature. Two decades later there is not a single viable nanomagnetic logic chip in sight, let alone one that is a commercial success. In this perspective article, I offer my reasons on why this has come to pass. I present a realistic and tempered vision of nanomagnetic logic, pointing out many misconceptions about this paradigm, flaws in some proposals that appeared in the literature, shortcomings, and likely pitfalls that might stymie progress in this field.

## I. NANOMAGNETIC BOOLEAN LOGIC

Traditional logic gates use a transistor as the primitive binary digital switch to implement the conditional dynamics of a gate. The transistor switch is fast, reliable, has isolation between the input and output terminals, has voltage or current gain, and also has a high conductance on/off ratio – all of which are conducive to implementing logic circuits. It's inherent drawback, however, is that it is a volatile device and hence the output bit of an end-of-chain logic gate must be transferred to a remote memory bank for safekeeping. That bit may then have to be fetched back from the memory bank to execute the next instruction set and subsequently returned to the bank for preservation. This transfer back and forth between the processor (gate) and the memory slows down the computation and is ultimately the most serious bottleneck in von-Neumann architectures. In contrast, a nanomagnetic logic gate is non-volatile (because it is implemented with nanomagnets) and hence the output bit can be stored in-situ in the gate, thus eliminating the need to transfer it to a memory bank. This feature lends itself to powerful non-von-Neumann architectures (e.g. "computing in memory") and myriad other applications. It is this property (non-volatility) that initially spurred interest in nanomagnetic logic. Furthermore, it was also conjectured that a nanomagnetic switch may be more energy-efficient than a transistor (although this is not always true) but this advantage could be offset by the slower switching speed. In fact, the energy-delay product of a nanomagnetic switch has never been convincingly shown to be significantly smaller than that of a transistor. The switching speed however is a moot issue since the speed of a computational task is determined more by the nature of the problem and the architecture of the computer than the working speed of its constituent elements. A case in point is the human brain versus a digital supercomputer. Neurons respond slowly, in time scales of milliseconds, and yet the human brain easily outperforms the best digital supercomputer (whose transistors switch in ~100 picoseconds or less) in such tasks as face recognition. Since magnetic switches may be amenable to superior architectures because of their inherent non-volatility, their slowness may be (at least partially) offset by their compatibility with such architectures. Moreover, there is now significant research in anti-ferromagnetic switches which switch much faster than ferromagnetic switches [1] and they are capable of reaching the switching speed of a modern day transistor.

A recent study explored the effect of slow device switching speeds in Boolean logic and used the memristor as an example [2]. A memristor is typically an order of magnitude slower than a magnetic switch. The study focused on ternary logic and concluded that slow switching remains an impediment, but can be tolerated for specific types of circuitry. Although slow switching is undeniably a disadvantage, it is not the primary reason why magnetic devices are not the preferred constituents of Boolean circuits. The primary reason is the lack of error resilience (unreliability) that magnetic devices suffer from and that precludes their application in state-of-the-art Boolean logic circuitry. I discuss that later in this perspective.

The basic idea of implementing a binary switch (the primitive element of a logic gate) with a nanomagnet is a simple one. Consider a nanomagnet shaped like an elliptical disk, as shown in Fig. 1. Because of the shape anisotropy energy in such a nanomagnet (caused by the elliptical shape), the magnetization can point only along the major axis of the ellipse (easy axis), either pointing to the left, or to the right, as shown in Fig. 1(a). If the nanomagnet has significant surface magnetic anisotropy, then the magnetization can point normal to the surface – either up or down – as shown in Fig. 1(b). The former kind of nanomagnets – of the type shown in Fig. 1(a) – are said to possess in-plane magnetic anisotropy (IPA) whereas those of the latter kind – shown in Fig. 1(b) – are said to possess perpendicular magnetic anisotropy (PMA). PMA nanomagnets are more immune to shape and size variations than IPA nanomagnets and hence preferred in many applications. In the ensuing discussion, we will focus on IPA nanomagnets, but we will return to PMA nanomagnets when the distinction between them is important.

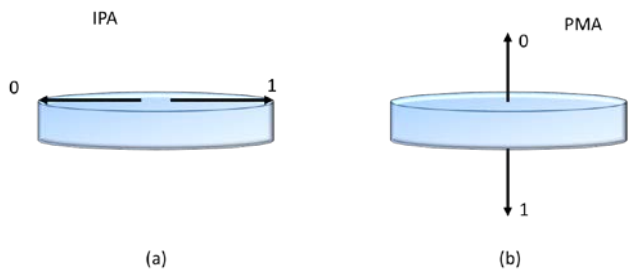

FIGURE 1. A nanomagnet shaped like an elliptical disk has two stable magnetization orientations which can encode the binary bits 0 and 1. (a) in-plane magnetic anisotropy, and (b) perpendicular magnetic anisotropy.

The two stable orientations of the magnetization shown in Fig. 1(a) can encode the binary bits 0 and 1. If we can switch the magnetization between these two orientations with an external agent, then we will realize the *primitive binary switch*. This is the basic idea of a nanomagnetic switch. The nanomagnet is of course assumed to be ideal, with no defect or pinning site that can give rise to metastable states, i.e. metastable orientations that deviate from the major axis of the ellipse. We will also assume that the magnetic material is amorphous and hence there is no magneto-crystalline anisotropy that could spawn additional stable or metastable states. Finally, we will assume that the nanomagnet is small enough to have a single domain, meaning that all the spins inside the nanomagnet point in the same direction because of mutual exchange interaction. When we flip the magnetization to switch the bit, all the spins in the single-domain nanomagnet *rotate in unison*, acting like one giant classical spin [3]. This is called coherent switching and it may reduce energy dissipation during the switching process.

The contrast with the transistor switch is immediately clear. In a transistor, the two bits 0 and 1 are encoded in two conductance states of the device (high and low). These two states are demarcated by the amount of electrical charge stored in the device. More charge present might make the device more conductive and less charge present will make it less conductive. Thus, the two bits are actually encoded in the amount of charge present in the device. To change the conductance state from one to the other (and thus switch the bit), electrical charge must flow into or out of the device [4], leading to the passage of a current and the inevitable associated energy dissipation. This is a shortcoming of all charge-based devices. After all, charge is a scalar quantity and has only magnitude and no direction. Hence if we use charge as the state variable to encode bit information, as we do in a transistor, then we must do so using two different *magnitudes* or amounts of charge. Switching from one binary bit to the other would then necessitate changing the amount of charge stored in the device, accompanied by unavoidable current flow and energy dissipation. This is the reason why charge-based devices tend to be relatively energy-inefficient. Of course, that begs the question if a nanomagnetic switch can be switched without causing any current flow and therefore be more energy-efficient. I discuss this next.

## II. DOES CURRENT FLOW HAVE TO ACCOMPANY THE SWITCHING OF A NANOMAGNET?

If we use the magnetization of a nanomagnet to encode the binary bits, as in Fig. 1, then we are using a vector quantity whose direction (not magnitude) represents the bit information. On the surface it seems that flipping the direction should not require any current flow (and indeed no current needs to flow inside the nanomagnet) which would eliminate any energy dissipation associated with current flow when we switch. This is actually not true. Flipping the magnetization can be achieved in many ways, but they typically *do require* flow of some current, if not within the nanomagnet itself, then somewhere else. We can think of flipping the magnetization with a local magnetic field, but that would require generating that field with a local current. We can also think of flipping it with a spin polarized current, as in spin-transfer-torque (STT) [5] or spin-orbit-torque (SOT) [6] based switching, but generating a spin current without a corresponding charge current flowing somewhere (not necessarily within the nanomagnet) is unusual and not known to this author. There are voltage controlled mechanisms of flipping magnetization, such as voltage controlled magnetic anisotropy (VCMA) [7] or voltage generated strain [8], but where there is voltage, there is also current. The two are related by Ohm's law. Of course, it can be argued that if the voltage is dropped over an infinite resistance, then the current through the resistor will be zero. Unfortunately, even an infinite resistance will have a capacitance in parallel (even if it is a parasitic one), which will be charged by the voltage (resulting in a charging current) and that action will dissipate energy as long as the charging time is finite. In fact, most voltage controlled mechanisms of switching the magnetization of a nanomagnet (e.g. VCMA or voltage-generated stain) ultimately involve charging a capacitor and there is a charging current associated with that process. Hence, it is very unlikely, if not impossible, that one can flip magnetization without causing an accompanying current flow *somewhere*. Thus, there will be energy dissipation during switching the magnetization of a nanomagnet associated with current flow. Whether that dissipation is larger than what is encountered in switching a transistor, or smaller, depends on the type of transistor and the magnet switching mechanism employed. There are many ways to switch the magnetization of a nanomagnet and some are more energy-efficient than others. However, what is important to realize is that switching a nanomagnet also requires current flow to take place somewhere, even though the nanomagnet is not a charge-based switch. It is a misconception that nanomagnets can be switched without requiring current flow. Therefore the nanomagnetic switch is *not inherently* more energy-efficient than the transistor.

### III. IS IT TRUE THAT A NANOMAGNET DOES NOT LEAK AND HENCE THE STANDBY POWER DISSIPATION IN A NANOMAGNETIC SWITCH IS ZERO?

It is well-known that a complementary-metal-oxide-semiconductor (CMOS) switch has a non-zero standby power dissipation because both the p-channel device and the n-channel device making up the CMOS conduct non-zero amount of current when they are off, i.e., they "leak". The standby power dissipation is entirely a consequence of the non-ideality of CMOS causing the leakage. An ideal CMOS should not leak and should conduct current only during switching, resulting in no standby power dissipation at all. It is often believed by proponents of nanomagnetic logic and switches that nanomagnetic switches need not suffer from standby power dissipation because nanomagnets do not "leak". Furthermore, it is sometimes claimed that this has to do with the fact that a transistor is a "volatile" switch (because stored charges leak away) while a nanomagnet is "non-volatile". Let us examine this issue in more detail.

Whether nanomagnetic switches leak or not depends on how they are utilized in nanomagnetic logic circuits. For the purpose of this discussion, I will first classify nanomagnetic logic into two basic types: one that uses nanomagnets deposited on an insulating substrate, which interact with each other via dipole coupling to elicit logic functionality, and another that uses a magnetic tunnel junction (MTJ) [discussed later in this section] as the central element of a Boolean logic gate. The first type is among the oldest and has been termed "magnetic quantum cellular automata" [9-11]. A schematic depiction (not representing any particular gate or circuit) is shown in Fig. 2. Here the nanomagnets indeed do *not* leak charges (ideally) since no current ever passes through them.

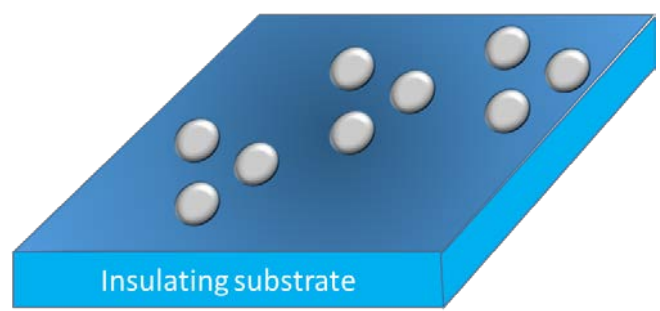

FIGURE 2. Nanomagnets deposited on an insulating substrate to implement magnetic quantum cellular automata type architecture. These nanomagnets do not have a leakage path and hence do not leak.

However, numerous groups have shown that this class is *too error-prone for logic* [12-17]. Hence the fact that there is no leakage is little consolation since the paradigm is too unreliable to be of any use in Boolean logic. I will revisit this paradigm later again in Section IV.

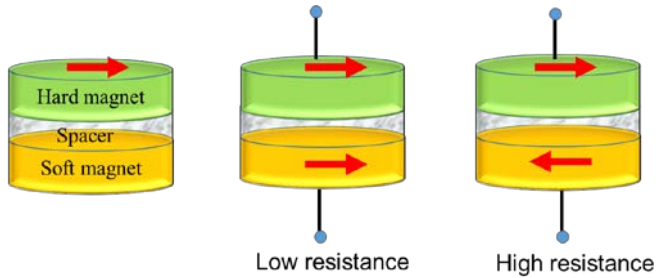

FIGURE 3. A magnetic tunnel junction (MTJ) converts magnetic orientation into electrical resistance and enables the electrical reading of a binary bit encoded in the magnetization orientation of a nanomagnet. It is therefore the most commonplace magnetic switch. Here we have shown an implementation with magnetic layers that have IPA, but a PMA implementation works similarly.

The other type uses a "magnetic tunnel junction" (MTJ) to function as a gate [18-21]. It is generally much more reliable than magnetic quantum cellular automata. The MTJ is shown in Fig. 3 and has been used as a memory cell, reader and writer of bit information stored in the magnetization states of nanomagnets, artificial neurons and synapses, Boltzmann machines, Bayesian inference engines, and Boolean logic gate. It has three layers shaped like elliptical disks. The two outer layers are ferromagnetic and the intervening spacer layer is an insulator. One of the ferromagnetic layers is a "hard" or "fixed" layer whose magnetization is permanently oriented along one of the two stable orientations. The other ferromagnetic layer (whose magnetization encodes the bit information) is the "soft" or "free" layer whose magnetization can point in either of the two stable directions and hence encode either bit 0 or bit 1. To "read" which bit has been encoded in the soft layer, one simply measures the electrical resistance between the two ferromagnetic layers. If the magnetizations of the two layers are mutually parallel, then the resistance is low and when they are antiparallel, the resistance is high. Since the magnetization of the hard layer is known and it is fixed, the measured resistance (high or low) will tell us whether the magnetization of the soft layer is pointing to the left or right. In other words, we can deduce whether the encoded bit is 0 or 1. This makes the MTJ an ideal "reader" of magnetic bit information. It can be also used to write bit information using STT, SOT, VCMA, etc. In fact, the MTJ can be used as a reader and writer in magnetic quantum cellular automata as well, although this is not always necessary.

In addition to the reading and writing functions, the MTJ can also be configured to act as a universal Boolean logic gate (NAND, NOR, etc.) [18-21]. In that role, it essentially acts as a magnetic switch with two conductance states – on and off. Note, however, that this is exactly similar to the case of a transistor switch where the two bits are encoded in two conductance states. Reading the bit requires flow of current (because we have to measure the conductance) and that is accompanied by energy dissipation. Writing the bit also requires flow of current somewhere (as discussed earlier) and causes energy dissipation. There is no reason to presuppose that this dissipation is always lower than that encountered in a transistor switch although there is also no fundamental reason why it cannot be lower.

Since we are discussing leakage, we will compare the ratio of the two conductances representing the two bit states in a transistor with that in a nanomagnetic switch like the MTJ. This is the so-called "on/off ratio" and it determines how leaky a switch is. For transistors, this value would typically exceed $10^5$. Consider a field-effect transistor with a threshold voltage of 0.5 V. Considering a constant sub-threshold slope of even 100 mV/decade (the minimum in an ordinary transistor at room temperature is about 60 mV/decade), the on/off ratio is $10^{(0.5/0.1)} = 10^5$. The highest on/off ratio reported in an MTJ so far is less than 10:1 at room temperature [22]. Therefore, for the same on-current, the off-current in the MTJ switch is at least four orders of magnitude *larger* than in the transistor switch, making the MTJ switch actually *leak much more* than the transistor switch.

When a memory cell is implemented with an MTJ, a transistor is usually placed in series with the latter and synchronized with it (i.e. the transistor is off when the MTJ is off) and this does reduce the leakage considerably, but MTJ based Boolean logic paradigms do not admit of such "series transistors". In fact, such a transistor may impair the functionality of the logic gate. Therefore, the MTJ based logic gates will leak much more than transistor based gates!

A clever strategy to ameliorate the high leakage associated with the low on/off ratios of MTJs was devised in *mLogic* [23] which used current steering to reduce the deleterious effect of leakage. However, it accomplished this with buffers and additional circuit overhead, which compromises circuit density, among other things.

## IV. UNRELIABILITY OF NANOMAGNETIC LOGIC

As stated earlier, the attractive feature of nanomagnetic Boolean logic is its *non-volatility*. Because a nanomagnetic gate is non-volatile, different stages of the computation need not be synchronized to a single clock and the gates can be operated asynchronously, leading, perhaps, to lower energy dissipation [18] regardless of whether the switch itself is particularly energy-efficient or not. It also enables easy implementation of non-von-Neumann architectures, as discussed earlier. Despite this attractive feature, nanomagnetic logic has not become mainstream and that may have happened because of its unreliability.

The switching error probability of a binary digital switch is the probability that an attempt to switch it from one stable state to the other fails owing to extraneous influences such as device defects or thermal noise. The switching error probability of a transistor is typically on the order of $10^{-15}$ [24], while the switching error probability of a nanomagnet is usually several orders of magnitude larger. The latter probability depends on the switching mechanism employed (e.g. spin-transfer-torque, spin-orbit torque, VCMA, strain, etc.) and therefore also on the amount of energy dissipated during switching since these different mechanism have different levels of energy dissipation. Generally, there is a trade-off between energy dissipation and error probability, which is true of both nanomagnets and transistors [4]. For comparable amounts of energy dissipation, the transistor is *much more error-resilient*, i.e. it has a much lower error probability. Strain-mediated switching of nanomagnets is one of the most energy-efficient nanomagnet switching mechanisms [25]. Simulations have shown that the room-temperature error probability in switching *pristine* defect-free IPA nanomagnets with strain exceeds $10^{-9}$ [12-15] and it is significantly higher if defects are present [26]. PMA nanomagnets are less vulnerable to structural variations (shape and size variations) but they are still very vulnerable to thermal noise whose effects are not ameliorated by PMA. This does not bode well for nanomagnetic logic. Logic has stringent requirements for reliability since bit errors in logic circuits propagate. If the output bit of a logic gate is erroneous and that is fed as the input to the next logic gate, then the output bit of that latter gate will also be corrupted and so on. In other words, errors are contagious in logic circuits. In contrast, errors in memory circuits are not contagious. If the bit stored in one memory cell is corrupted, then it does not corrupt any other cell. Errors are also dynamic in logic chips and relatively static in memory chips. That is why many on-chip error correction schemes exist for memory, but not for logic.

The fact that magnetic switches are unsuitable for logic has been pointed out by a number of authors [26, 27]. One particularly vulnerable paradigm is the so-called "magnetic quantum cellular automata" (mentioned earlier) which relies on dipole coupling between neighboring nanomagnets to implement Boolean logic gates [9-11]. The high error rates in this paradigm accrue from the fact that dipole coupling is usually too weak to withstand thermal noise. Dipole coupling based circuitry is also not scalable in size since the dipole coupling energy is proportional to the square of the volume of the nanomagnets. Our own experiments with *ultra-low energy* dipole-coupled NOT gates implemented with strain-switched nanomagnets (strain is generated with voltage) have shown that the error probabilities are several tens of percent when the energy dissipated to switch is a few to few tens of aJ [28]. Of course, there is always a trade-off between energy-cost and reliability [4]. These experiments suggest that *low-energy* dipole coupled architectures may not be viable for logic after all.

One claimed experimental demonstration of a dipole coupled majority logic gate using IPA nanomagnets reported the gate error probability to be 75% [10]! This is really no surprise since, among other things, IPA nanomagnet-based architectures are extremely sensitive to even slight lateral misalignment of the nanomagnets if global clocking mechanisms are used and this makes the paradigm extremely error-prone and unrealistic for logic applications [29]. A later experiment used PMA nanomagnets to construct the majority logic gate with the hope that the reliability will improve because PMA nanomagnets' operations are relatively immune to shape and size variations as well as, perhaps, misalignment [11]. However, even though a low error rate was claimed, it was not mentioned what this error rate was (probably because insufficient data were available). Another paper claimed "error-free" propagation of bits in a chain of nanomagnets making up an inverter, using the same paradigm [30], but there is no mention of how many times the propagation was attempted and how times it succeeded, or if more than one chain was probed. This then provides no information on the error probability. If the construct worked 1 out of 1 time, it does not tell us whether it also works 10 out of 10 times or 3 out of 10 times, etc. All we can infer from this experiment is that the error probability is not 100%, but then it could be anything between 0% and 99.999…%.

In 1956, von-Neumann had shown that the maximum tolerable error probability in a *single* majority logic gate working in isolation is 0.0073 [31]. The error probability will have to be several orders of magnitude smaller than that if the gate has to work in a "circuit" comprising millions of gates. Suffice it to say then that constructs like magnetic quantum cellular automata are too unreliable to harness for Boolean logic operations. Yet, numerous theoretical proposals for shift registers, full adders and more complex circuits that utilize the notion of magnetic quantum cellular automata continue to appear in the scientific literature and will probably continue in the future. Invariably they assume millions of pristine defect-free nanomagnets all working at a temperature of 0 K (when no thermal noise is present) and all of them switch with 0% error probability. No thought is given to the fact that not a single magnetic quantum cellular automata system (that performs a useful calculation) exists anywhere in the world, even two decades after the first proposal appeared. There has been no industrial interest in this idea in 20 years which could be ascribed to its non-viability.

The MTJ based proposals are much more reliable than dipole coupled architectures like magnetic quantum cellular automata because they do not rely on weak dipole coupling, but they have their own disadvantages as well. One of them, termed "all-spin logic" because it obviates the need for spin-to-charge conversion at any stage [19], relies on complicated clocking schemes for operation. The error probability of this paradigm has not been reported, but it is likely to be high because of the need for complicated clocking. Another [20] uses an MTJ to implement a full adder in a hybrid circuit configuration. Its error-resilience is unknown. Yet another [21] uses a basic strain switched magnetic tunnel junction to implement a universal Boolean logic gate and reports a theoretical error probability of $\sim 10^{-8}$ at room temperature. While these proposals are interesting and based on sound science, their reliability remains questionable, not to mention the fact that the poor on/off ratio leads to other undesirable effects. However, they may be relatively energy-efficient and exhibit a reasonably small energy-delay product.

There is, unfortunately, almost no prior study of the reliability of MTJ-based nanomagnetic logic. Ref. [32] undertook such a study using a 4-terminal MTJ based logic gate termed "m-Gate" in the spirit of m-Logic [23]. These gates are built with current-driven MTJs switched via spin transfer torque. The authors showed that the m-Gate is more reliable than m-Logic and calculated the error probabilities of 2 and 3-terminal NAND and NOR gates as a function of the power supply voltage. They showed that the required gate error probability of $10^{-15}$ can be obtained when the power supply voltage is 40-50 V! At a reasonable power supply voltage of 1 V, the error probability is almost 100%! This clearly calls into question the viability of such MTJ-based nanomagnetic logic.

Despite all their shortcomings for logic, MTJs are of course very suitable for "memory" as opposed to "logic". Memory is much more forgiving of errors than logic because if a single cell is corrupted, it does not "infect" any other cell. Moreover, the high leakage in the MTJ (because of the poor on/off ratio) is mitigated by placing a transistor in series with the MTJ in a memory cell and that transistor is used for reading and writing. MTJ based memory has become mainstream because of the excellent endurance, high density, very long retention times, good latency, and compatibility with a crossbar architecture. Two terminal memory is usually implemented by writing bit information with spin-transfer torque, while three-terminal memory (with much lower write energy dissipation but larger footprint) is implemented by writing data using spin-orbit torque. Although this perspective addresses magnetic logic and not magnetic memory, we briefly digress into this discussion here for the sake of completeness.

Another genre of nanomagnetic logic proposals that is popular utilizes magnetic domain walls and encodes bit information in the magnetization direction of domains [33-35]. They do not suffer from the problem of leakage or poor on/off ratio because they do not employ MTJs in the logic processing itself, but their error resilience is not likely to be high since domain wall magnetization can be vulnerable to stray magnetic fields and thermal noise. Many proposals require the domains to propagate through racetracks and domains may get trapped or stuck at locations within the racetrack that have defects, resulting in error. Therefore, their suitability for logic is again questionable.

## V. NANOMAGNETIC LOGIC PROPOSALS THAT DO NOT SATISFY THE REQUIREMENT OF CONCATENABILTY

It is well-known among logic designers and electrical engineers that logic computing has to fulfill certain fundamental requirements in *addition* to performing Boolean operations (see, for example, refs. [19, 21, 36]). Among them are: concatenability, non-linearity, isolation between input and output, gain, universality, reliability and scalability. Concatenability implies that the output of one logic gate can be fed *directly* to the input of the next logic gate to implement combinational and sequential circuits for performing logic computing. Obviously, this would require that the input and output variables (bits) be encoded in the *same* physical quantity. For example, if the input is encoded in voltage states, then the output must also be encoded in voltage states in order to be fed directly to the input(s) of the next logic gate. If the input(s) and output(s) are dissimilar quantities (e.g. the input bit is encoded in voltage and the output bit is encoded in light intensity), then *transducers* will be required to precede every input port in order to transduce the arriving output signal from the preceding gate to the correct physical form that the input terminal of the successor gate can accept. Converting light intensity to current, for example, will require a photodetector as the transducer.

I illustrate this with a simple cartoon in Fig. 4 using a two-input and one-output gate like the traditional AND, NAND, OR or NOR. Suppose that one of the inputs is a "triangle", the other is a "square", while the output is a "circle" – all dissimilar shapes. Then the output of first logic gate can never be fed directly to either input of the second since one input of the second gate can accept only a "triangle" and the other a "square" and neither can accept a "circle". Here, one would require transducers to (metaphorically) transduce the "circle" to a "triangle" and a "square" at every logic stage. These transducers may be so costly in terms of hardware (device footprint, energy dissipation, cost, etc.) that it will not be worth it. That is why concatenability is a fundamental requirement for logic computing [19, 21, 35] because it eliminates the need for any transducer.

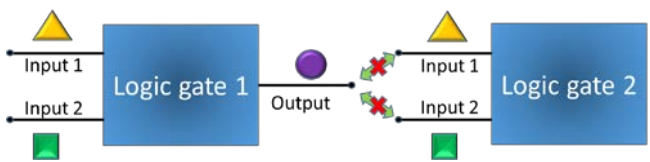

FIGURE 4. Cartoon to illustrate the requirement of "concatenability" in logic circuits.

Sometimes, the transducer may be simple. For example, if the output variable is voltage and one input variable is current, then a simple resistor can act as the transducer. However, in many cases, the transducer may have to be much more complex. In fact, in some logic proposals that I will discuss next, the transducer will have to be so complex that it will consume much more energy than the gate itself, its footprint will be much larger than that of the gate, and its operation may be slower than that of the gate. The performance is then dictated by the transducer and not the gate! This is the reason why concatenability is a fundamental requirement of logic circuits and those that violate it are simply impractical.

Many published magnetic logic proposals actually violate this requirement. One of the earliest ones among them is the "magnetic quantum cellular automata" proposal of ref. [9]. This deals with 2-input and 1-output Boolean logic operations like AND, NAND, NOR, and OR that are carried out with dipole coupled nanomagnets. In this scheme, one input bit is encoded in the sign (positive or negative) of a magnetic field pulse, another is encoded in the phase of an oscillating field and the output is encoded in the magnetization state of a chain of magnetic dots measured with the magneto-optical Kerr effect. They are all *dissimilar quantities*! Therefore, immensely complex transducers will be needed preceding every gate to convert the magnetization of a chain of nanomagnets into both the *sign* of a magnetic pulse and the *phase* of an oscillating field. This is not a practical paradigm.

Another non-concatenable logic proposal is based on an MTJ-like device acting as the primitive switch to implement Boolean logic functions [18]. Here the two input bits are directions (or polarities) of a current and the output bit is the magnetization of a layer. The magnetization can be converted to an electrical resistance with an MTJ-like device and then the electrical resistance can be converted into a current *magnitude* with a constant voltage source. Finally, the current magnitude has to be converted into a current polarity (or direction) with some kind of a current level shifter. Suffice it to say that this assembly of transducers that must precede every gate will detract seriously from the efficacy of such a paradigm.

Yet another non-concatenable scheme has appeared more recently. Ref. [35] has proposed implementing Boolean logic gates with magnetic elements using current density (to induce spin-orbit torque) as one input, laser fluence as the other input and anomalous Hall effect voltage as the output. An electro-optic modulator type device will be needed as a transducer to convert the voltage into laser fluence and a resistor will be needed as the other transducer to convert the voltage into current density, and they will be required at the input terminals of every logic gate. It does not matter how good the electro-optic modulator is. Its very presence is a nuisance and the truth is that *it should not be required at all*. Needless to say, this scheme is not practical either. Unfortunately, schemes like this continue to appear in many applied physics journals and will probably continue to appear in future.

The non-concatenability flaw may not be unique to nanomagnetic logic gates and my list of (proposed) magnetic logic gates that suffer from this flaw (that I have mentioned above) may not be exhaustive either. The purpose of this discussion is to merely show that many nanomagnetic logic proposals are in violation of the basic tenets of logic design. Their claims of superior energy efficiency, smaller footprint, etc. (if they are made) are meaningless because the energy dissipation and the footprints of the complex transducers have not been factored in. In almost all of these proposals, the need for the transducers (arising out of the failure to satisfy the concatenability requirement) was not even realized.

There are some magnetic logic proposals that recognize the need for concatenability and avoid using transducers at every stage, but they require additional overhead such as a reset scheme and domino-style clocking [37], which would be inconvenient albeit not as debilitating as requiring complex transducers preceding every gate. Suffice is to say that cascading magnetic logic gates often poses additional challenges that detract from their efficacy as Boolean logic processors.

## VI. CONCLUSIONS

New ideas naturally generate excitement and are sometimes embraced without sufficient scrutiny. Nanomagnetic (Boolean) logic ideas have been around for at least two decades and yet not a single one of them has become mainstream or commercially viable, despite all the ebullient claims of superiority and the undeniable benefit of non-volatility. Perhaps the primary reason for failure is the lack of error-resilience. Additionally, some proposals lack some essential features like concatenability and would therefore never be practical or useful. The science behind these proposals is usually sound, but many of these proposals flout basic principles of logic design, which has prevented them from coming to fruition. Furthermore, most purported experimental demonstrations of nanomagnetic logic, available in the literature, have demonstrated the operation of a *single gate* which is a far cry from a logic circuit where multiple gates have to be concatenated to elicit logic functionality. A single gate does not establish viability.

There are many nanomagnetic logic proposals in the literature. The purpose of this perspective is not to critique them all, but rather to point out potential pitfalls with some specific examples. There are daunting obstacles that stand in the way of implementing practical nanomagnetic logic schemes, some of which have been discussed here.

A serious research effort is currently underway to implement *non-Boolean* computing machinery with nanomagnets [38-51]. Their requirements are very different from those of Boolean logic and fortunately are well suited to the features of nanomagnetic switches, especially their attribute of non-volatility. These paradigms are relatively forgiving of switching errors, poor on/off ratio, and the slow switching speed of nanomagnetic switches. Many of these constructs actually use MTJs to realize such functions as quantum annealers [38], neurons and synapses [38-40], probabilistic computing modules [41], image processing (for encoding pixel color – black or white - in the conductance state of an MTJ) [42], Bayesian belief networks where the parent and child nodes are implemented with MTJs [43-45], restricted Boltzmann machines for image classification and recognition of handwritten digits [46], integer factorization [47], invertible logic [48], computer vision [49] and physically unclonable function (PUF) generation [50]. Low barrier nanomagnets, where the thermal stability is intentionally reduced allowing the magnetization to fluctuate randomly in time are very useful for binary stochastic neurons [51]. Many of these applications, e.g. neuromorphic processors are extremely forgiving of errors since they compute in a collective mode where the cooperative activities of many MTJ neurons acting in unison produce the result of the computation and the overall operation is not impaired if some of the neurons misfire or are defective. This is analogous to the human brain where the cognitive ability is not inhibited even if some neurons die or are erratic. This area may be the one where nanomagnets will (hopefully) find a niche and even steal a march over their transistor counterpart. Boolean logic is not the arena where nanomagnetic switches have, at least so far, made any significant impact.

## REFERENCES

1. P. A. Dowben, D. E. Nikonov, A. Marshall and Ch. Binek, " Magneto-electric antiferromagnetic spin-orbit logic devices", *Appl. Phys. Lett.*, vol. 116, no. 8, Art. No. 080502 (2020).
2. X. Wang, P. Zhou, J. K. Eshraghian, C. Y. Lin, H. H. C Iu, T-C Chang, and S-M Kang, "High-density memristor-CMOS ternary logic family", *IEEE Trans. Circ. Syst. I*, Digital Object Identifier 10.1109/TCSI.2020.3027693.
3. R. P. Cowburn, D. K. Koltsov, A. O. Adeyeye, M. E. Welland and D. M. Tricker, "Single-domain circular nanomagnets", *Phys. Rev. Lett.*, vol. 83, No. 5, pp. 1042-1045 (1999).
4. S. Bandyopadhyay, "Straintronics: Digital and analog electronics with strain-switched nanomagnets", *IEEE Open Journal of Nanotechnology*, vol, 1, pp. 570-64, Aug. (2020).
5. D. C. Ralph and M. D. Stiles, "Spin transfer torques", *J. Magn. Magn. Mater.*, vol. 320, No. 7, pp. 1190-1216 (2008).
6. I. M. Miron, et al., "Perpendicular switching of a single ferromagnetic layer induced by in-plane current injection", *Nature*, vol. 476, pp. 189-193 (2011).
7. P. K. Amiri and K. L. Wang, "Voltage-controlled magnetic anisotropy in spintronic devices", *SPIN*, vol, 2, Art. No. 1240002 (2012).
8. K. Roy, S. Bandyopadhyay and J. Atulasimha, "Energy dissipation and switching delay in stress-induced switching of multiferroic nanomagnets in the presence of thermal fluctuations", *J. Appl. Phys.*, vol. 112, No. 2, Art. No. 023914 (2012).
9. R. P. Cowburn and M. E. Welland, "Room temperature magnetic quantum cellular automata", *Science*, vol. 287, Issue 5457, pp. 1466-1468 (2000).
10. A. Imre, et al., "Majority logic gate for magnetic quantum dot cellular automata", *Science*, vol. 311, Issue 5758, pp. 205-208 (2006).
11. S. Breitkruetz, et al., "Majority gate for nanomagnetic logic with perpendicular magnetic anisotropy", *IEEE Trans. Magn.*, vol. 48, No. 11, pp. 4336-4339 (2012).
12. M. Salehi-Fashami, K. Munira, S, Bandyopadhyay, A. Ghosh and J. Atulasimha, "Switching of dipole-coupled multiferroic nanomagnets in the presence of thermal noise: Reliability of nanomagnetic logic", *IEEE Trans. Nanotechnol.*, vol. 12, No. 6, pp. 1206-1212 (2013).

13. M. M. Al-Rashid, S. Bandyopadhyay and J. Atulasimha, "Effect of nanomagnet geometry on reliability, energy dissipation and clock speed in strain clocked DC-NML", *IEEE Trans. Elec. Dev*., vol. 62, No. 9, pp. 2978-2986 (2015).
14. K. Munira, Y. Xie, S. Nadir, M. B. Forgues, M. Salehi-Fashami, J. Atulasimha, S. Bandyopadhyay and A. W. Ghosh, "Reducing error rates in straintronic multiferroic nanomagnetic logic by pulse shaping", *Nanotechnology*, vol. 26, No. 24, Art. No. 245202 (2015).
15. M. M. Al-Rashid, S. Bandyopadhyay and J. Atulasimha, "Dynamic error in strain-induced magnetization reversal of nanomagnets due to incoherent switching and formation of metastable states: A size-dependent study", *IEEE Trans. Elec. Dev*., vol. 63, No. 8, pp. 3307-3313 (2016).
16. F. M. Spedalieri, A. P. Jacob, D. E. Nikonov and V. P. Roychowdhury, "Performance of magnetic quantum cellular automata and limitations due to thermal noise", *IEEE Trans. Nanotechnol.*, vol. 10, No. 5, pp. 537-546 (2010).
17. D. B. Carlton, B. Lambson, A. Scholl, A. T. Young, P. D. Ashby, S. Dhuey and J. Bokor, "Investigation of defects and errors in nanomagnetic logic circuits", *IEEE Trans. Nanotechnol.*, vol. 11, No. 4, pp. 760-762 (2012).
18. A. Ney, C. Pampuch, R. Koch and K. Ploog, "Programmable computing with a single magnetoresistive element", *Nature*, vol. 425, pp. 485-487 (2003).
19. B. Behin-Aein, D. Datta, S. Salahuddin and S. Datta, "Proposal for all-spin logic device with built-in memory", *Nature Nanotechnol*., vol. 5, pp. 266-270 (2010).
20. S. Matsunaga, et al., "Fabrication of non-volatile full adder based on logic-in-memory architecture using magnetic tunnel junctions", *Appl. Phys. Express*, vol. 1, No. 9, Art. No. 091301 (2008).
21. A. K. Biswas, J. Atulasimha and S. Bandyopadhyay, "An error-resilient non-volatile magneto-elastic universal logic gate with ultralow energy-delay product", *Sci. Rep*., vol. 4, Art. No. 7553 (2014).
22. S. Ikeda, J. Hayakawa, Y. Ashizawa, Y.M. Lee, K. Miura, H. Hasegawa, M. Tsunoda, F. Matsukura and H. Ohno, "Tunnel magnetoresistance of 604% at 300 K by suppression of Ta diffusion in CoFeB/MgO/CoFeB pseudo-spin-valves annealed at high temperature", *Appl. Phys. Lett*., vol. 93, No. 8, Art. No. 082508 (2008).
23. D. Morris, D. Bromberg, J-G Zhou and L. Pileggi, "mLogic: Ultra-low voltage non-volatile logic circuits using STT-MTJ devices", Proceedings of the 49th Annual Design Automation Conference, June 2012, pp. 486-491. https://doi.org/10.1145/2228360.2228446
24. A. D. Patil, S. Manipatruni, D. Nikonov, I. A. Young and N. R. Shanbhag, "Shannon-inspired statistical computing to enable spintronics", arXiv: 1702.06119.
25. N. D'Souza, et al., "Energy-efficient switching of nanomagnets for computing: Straintronics and other methodologies", *Nanotechnology*, vol. 29, No. 44, Art. No. 442001 (2018).
26. D. Winters, M. A. Abeed, S. Sahoo, A. Barman, and S. Bandyopadhyay, "Reliability of magneto-elastic switching of non-ideal nanomagnets with defects: A case study for the viability of straintronic logic and memory", *Phys. Rev. Appl*., vol. 12, No. 3, Art. No. 034010 (2019).
27. A. D. Patil, S. Manipatruni, D. E. Nikonov, I. A. Young and N. R. Shanbhag, "Error-resilient spintronics via the Shannon-inspired model of computation", *IEEE J. Explor. Solid State Comput. Dev. and Circ*., vol. 5, No. 1, pp. 10-18 (2019).
28. H. Ahmad, J. Atulasimha and S. Bandyopadhyay, "Electric field control of magnetic states in isolated and dipole-coupled FeGa nanomagnets delineated on a PMN-PT substrate" *Nanotechnology*, vol. 26, No. 40, Art. No. 401001 (2015).
29. S. Bandyopadhyay and M. Cahay, "Electron spin for classical information processing: A brief survey of spin-based logic devices, gates and circuits", *Nanotechnology*, vol. 20, No. 41, Art. No. 412001 (2011).
30. I. Eichwald, A. Bartel, J. Kiermaier, S. Breitkruetz, G. Csaba, D. Schmitt-Landsiedel and M. Becherer, "Nanomagnetic logic: Error-free, directed signal transmission by an inverter chain", *IEEE Trans. Magn*., vol. 48, No. 11, pp. 4336-4339 (2012).
31. J. von Neumann, "Probabilistic logic and the synthesis of reliable organisms from unreliable components", *Autom. Stud*., vol. 34, pp. 43-98 (1956).
32. V. Jamshidi and M. Fazeli, "Pure magnetic logic circuits: A reliability analysis", *IEEE Trans. Magn.*, vol. 54, No. 10, Art. No. 3401010 (2018).
33. D. A. Allwood, et al., "Magnetic domain wall logic", *Science*, vol. 309, Issue 5741, pp. 1688-1692 (2005).
34. Z. Luo, et al., "Current-driven magnetic domain wall logic", *Nature*, vol. 579, pp. 214-218 (2020).
35. B. Zhang, D. Zhu, Y. Xu, X. Lin, M. Hehn, G. Malinowski, W. Zhao and S. Mangin, "Optoelectronic domain wall motion for logic computing", *Appl. Phys. Lett.*, vol. 116, No. 25, Art. No. 252403 (2020).
36. R. Waser (ed). *Nanoelectronics and Information Technology*, Ch. III, (Wiley-VCH, 2003).
37. A. Jaiswal and K. Roy, "MESL: Proposal for non-volatile cascadable magneto-electric spin logic", *Sci. Rep*., vol. 7, Art. No. 39793 (2017).
38. J. Grollier, D. Querlioz, K. Y. Camsari, K. Evershcor-Sitte, S. Fukami and M. D. Stiles, "Neuromorphic spintronics", *Nature Electron*., https://www.nature.com/articles/s41928-019-0360-9.
39. A. K. Biswas, J. Atulasimha and S. Bandyopadhyay, "Straintronic spin neuron", *Nanotechnology*, vol. 26, No. 28, Art. No. 285201 (2015).
40. C. Cui, O. G. Akinola, N. Hassan, C. H. Bennett, M. J. Marinella, J. S. Friedman and J-A C. Incorvia, "Maximized lateral inhibition in paired magnetic domain wall racetracks for neuromorphic computing", *Nanotechnology*, vol. 31, Art. No. 294001 (2020). https://doi.org/10.1088/1361-6528/ab86e8.
41. K. Y. Camsari, B. M Sutton and S. Datta, "p-bits for probabilistic spin logic", *Appl. Phys. Rev*., vol. 6, No. 1, Art. No. 011305 (2019).
42. M. A. Abeed, A. K. Biswas, M. M. Al-Rashid, J. Atulasimha and S. Bandyopadhyay, "Image processing with dipole coupled nanomagnets: Noise suppression and edge enhancement detection", *IEEE Trans. Elec. Dev*., vol. 64, No. 5, pp. 2417-2424 (2017).
43. S. Khasanvis, M. Li, M. Rahman, M. Salehi-Fashami, A. K. Biswas, J. Atulasimha, S. Bandyopadhyay and C. A. Moritz, "Self-similar magnetoelectric nanocircuit technology for probabilistic engines", *IEEE Trans. Nanotechnol.*, vol. 14, No. 6, pp. 980-991 (2015).
44. S. Nasrin, J. L. Drobitch, P. Shukla, T. Tulabandhula, S. Bandyopadhyay and A. R. Trivedi, "Bayesian reasoning machine on a magneto-tunneling junction network", *Nanotechnology*, vol. 31, Art. No. 484001 (2020). https://doi.org/10.1088/1361-6528/abae97.
45. M. T. McCray, M. A. Abeed and S. Bandyopadhyay, "Electrically programmable probabilistic bit anti-correlator on a nanomagnetic platform", *Sci. Rep*., vol. 10, Art. No. 12361 (2020).
46. S. Nasrin, J. L. Drobitch, S. Bandyopadhyay, and A. R. Trivedi, "Low power restricted Boltzmann machine using mixed-mode magneto-tunneling junctions", *IEEE Elect. Devices Lett.*, vol. 40, No. 2, pp. 345-348 (2019).
47. W. A. Borders, A. Z. Pervaiz, S. Kukami, K. Y. Camsari, H. Ohno and S. Datta, "Integer factorization using stochastic magnetic tunnel junction", *Nature*, vol. 573, pp. 390-393 (2019).

48. A. Z. Pervaiz, L. A. Ghantasala, K. Y. Camsari and S. Datta, "Hardware emulation of stochastic p-bits for invertible logic", *Sci. Rep.*, vol. 7, Art. No. 10994 (2017).
49. S. Bhanja, D. K. Karunaratne, R. Panchumarthy, S. Rajaram and S. Sarkar, "Non-Boolean computing with nanomagnets for computer vision applications", *Nature Nanotechnol.*, vol. 11, pp. 177-183 (2016).
50. G. Finocchio, T. Moriyama, R. De Rose, G. Siracusano, M. Lanuzza, V. Puliafito, S. Chiappini, F, Crupi, Z. Zeng, T. Ono and M. Carpentieri, "Spin-orbit torque based physical unclonable function", *J. Appl. Phys.*, vol. 128, No. 3, Art. No. 033904 (2020).
51. O. Hassan, R. Faria, K. Y. Camsari, J. Z. Sun and S. Datta, "Low barrier magnet design for efficient hardware binary stochastic neurons", *IEEE Mag. Lett.*, vol. 10, Art. No. 4502805, May (2019).